\documentclass{ptptex}
\usepackage{wrapft}
\usepackage[pdftex]{graphicx}

\begin{document}

\markboth{A. Kagawa, M. Morimoto, et.al.
}
{Non-zero tensor condensates in cold quark matter
}


\newcommand{\bbra}[1]{\langle\!\langle {#1} |}     
\newcommand{\kket}[1]{| {#1} \rangle\!\rangle}     
\newcommand{\rbra}[1]{( {#1} |}     
\newcommand{\rket}[1]{| {#1} )}     
\newcommand{\rdket}[1]{|\!| {#1} )}     
\newcommand{\rrket}[1]{| {#1} ))}     
\newcommand{\dbra}[1]{\langle {#1} |\!|}     
\newcommand{\rdbra}[1]{( {#1} |\!|}     
\newcommand{\dket}[1]{|\!| {#1} \rangle}     
\newcommand{\maru}[1]{\breve{#1}} 
\newcommand{\wtilde}[1]{\widetilde{#1}} 
\newcommand{\lsim}{{\stackrel{<}{\sim}}}
\newcommand{\gsim}{{\stackrel{\displaystyle >}{\raisebox{-1ex}{$\sim$}}}}
\newcommand{\lal}{\langle\!\langle}
\newcommand{\rar}{\rangle\!\rangle}
\newcommand{\wb}[1]{\overline{#1}}
\newcommand{\vect}[1]{\overrightarrow{#1}}
\newcommand{\ovl}[1]{\overline{#1}}
\newcommand{\braket}[1]{\langle{#1}\rangle}
\def\beq{\begin{eqnarray}}
\def\eeq{\end{eqnarray}}
\def\bsub{\begin{subequations}}
\def\esub{\end{subequations}}
\def\beq{\begin{eqnarray}}
\def\eeq{\end{eqnarray}}
\def\bsub{\begin{subequations}}
\def\esub{\end{subequations}}
\def\b{\begin{equation}}
\def\bs{\begin{split}}
\def\es{\end{split}}
\def\e{\end{equation}}

\title{
Non-zero tensor condensates in cold quark matter within the three-flavor Nambu-Jona-Lasinio model with 
the Kobayashi-Maskawa-'t Hooft interaction
}

\author{Ai {\sc Kagawa}$^1$, Masatoshi {\sc Morimoto}$^1$, Yasuhiko {\sc Tsue}$^{2,3}$, {Jo\~ao da {\sc Provid\^encia}}$^{3}$,\\
{Constan\c{c}a {\sc Provid\^encia}}$^{3}$ 
and {Masatoshi {\sc Yamamura}}$^{3,4}$
}

\inst{$^{1}${Graduate School of Integrated Arts and Science, Kochi University, Kochi 780-8520, Japan}\\
$^{2}${Department of Mathematics and Physics, Kochi University, Kochi 780-8520, Japan}\\
$^{3}${CFisUC, Departamento de F\'{i}sica, Universidade de Coimbra, 3004-516 Coimbra, Portugal}\\
$^{4}${Department of Pure and Applied Physics, 
Faculty of Engineering Science,\\
Kansai University, Suita 564-8680, Japan}\\
}

\maketitle


\abst
The possible formation of tensor condensates originated from a tensor-type interaction between quarks 
is investigated in the three-flavor Nambu-Jona-Lasinio model including the 
Kobayashi-Maskawa-'t Hooft interaction, which leads to flavor mixing. 
It is shown that independent two tensor condensates appear 
and a tensor condensate related to the strange quark 
easily occurs by the effect of the flavor mixing compared with one related to light quarks.  
Also, it is shown that 
the tensor condensate related to the strange quark 
appears at a slightly smaller chemical potential if the
Kobayashi-Maskawa-'t Hooft interaction is included,
due to the flavor mixing effect.
It is also shown that the two kinds of tensor condensates may  coexist
in a certain quark chemical potential due to the flavor mixing.



\section{Introduction}

One of recent interests in many-quark system governed by 
quantum chromodynamics (QCD) is to clarify the existence of 
various phases on the plane spanned by the quark chemical potential and temperature.\cite{FH} 
As is indicated by many authors, 
there may exist various phases such as the color superconducting phase 
\cite{ARW,IB,CFL}, the quarkyonic phase,\cite{McL} 
the inhomogeneous chiral condensed phase,\cite{NT}
the quark ferromagnetic phase,\cite{Tatsumi} the color-ferromagnetic phase,\cite{Iwazaki}
the spin polarized phase due to the axial vector interaction\cite{NMT,TMN,Morimoto,Morimoto2} 
or due to the tensor interaction
\cite{BJ,IJMP,oursPTP,oursPTEP1,oursPTEP2,oursPTEP3,oursPTEP4,oursPR,Ferrer,MT,MNYY}.

In order to investigate the phase structure in quark matter at finite baryon density, 
various effective models 
of QCD are used because in the region of large quark chemical potential, 
the numerical simulation by using the lattice QCD did not work until now. 
One of the effective models of QCD, the Nambu-Jona-Lasinio (NJL) model \cite{NJL} 
is widely used \cite{Klevansky,HK}
because it contains  chiral symmetry, an important QCD symmetry. 
This model has been used to describe quark matter in the region with 
large quark chemical potential at low temperature\cite{Buballa}.
The extended NJL model, which includes a tensor-type four-point interaction 
and/or the vector-pseudovector-type four-point interaction between quarks, is 
introduced to investigate the possible formation of a tensor condensate and/or pseudovector condensate, 
which may lead to the quark spin polarization. 
\cite{BJ,IJMP,oursPTP,oursPTEP1,oursPTEP2,oursPTEP3,Ferrer,MT,MNYY,Maedan} 
If the quark spin polarization gives rise to spontaneous 
magnetization in quark matter, it may give origin to the 
strong magnetic field of compact stars such as neutron stars and magnetars.
\cite{HL}

In this paper, we concentrate on the spin polarization due to the tensor condensate originated 
from the tensor interaction between quarks in an 
extended NJL model. 
In our previous paper in \citen{oursPTEP4,oursPR,oursPR2}, the spin polarization 
due to the tensor condensate has been investigated widely in the case of the 
two-flavor NJL model. 
In this paper, we extend our previous work to the case of three-flavor NJL model. 
Previously, we have examined a possibility of conflict and/or coexistence of both the tensor condensate and 
the color-flavor locking condensate \cite{oursPTEP2}. 
In three-flavor case, we have shown that two typical tensor condensates, which we name as 
$F_3\ (=\langle {\bar \psi}\Sigma_3\lambda_3 \psi\rangle)$ and 
$F_8\ (=\langle {\bar \psi}\Sigma_3\lambda_8 \psi\rangle)$, appear,
where $\Sigma_3$ and $\lambda_a$ are, respectively, the
spin matrix with the third component ($z$-component) and the Gell-Mann
matrices for $a=1,...,8$ or the identity matrice times $\sqrt{2/3}$ ($a=0$). 
Also, a conjecture that $F_8=F_3/\sqrt{3}$ is satisfied was given.
However, in that paper, we have neglected both the flavor-symmetry breaking and the flavor mixing. 
If the flavor-symmetry is broken, the relation $F_8=F_3/\sqrt{3}$ is not valid.

In a recent paper\cite{India}, a spin polarization due to a tensor-type interaction in $2+1$ flavor 
NJL model has been investigated. 
The authors have introduced the two tensor condensates $F_3$ and $F_8$ correctly. 
However, they have assumed $F_8=F_3/\sqrt{3}$ in most part of their analysis for numerical simplicity, 
while they have treated $F_8$ and $F_3$ independently only in the final part of their paper.
As was previously mentioned, the approximate condition
$F_8=F_3/\sqrt{3}$ is valid  in the chiral limit, i.e.
if all the current quark masses are zero. 
Thus, we reanalyze the possible formation of the tensor condensates in 3-flavor cold quark matter at finite density. 
In three-flavor case, it is well known that the quark-flavor mixing occurs through 
the six-point interaction between quarks in the NJL model. 
This interaction is called the Kobayashi-Maskawa-'t Hooft interaction or 
the determinant interaction.
\cite{KM,determinant} 
Thus, in this paper, we especially focus our attention on the effects of the flavor symmetry breaking and 
the flavor mixing through 
the Kobayashi-Maskawa-'t Hooft interaction 
on the tensor condensates.

This paper is organized as follows: 
In the next section, the mean field approximation for the NJL model 
with tensor-type four-point interaction between quarks 
is given. Then, both the quark and antiquark condensate, namely chiral condensate, 
and the two-type tensor condensates are introduced 
and further, the thermodynamic potential is evaluated at zero temperature 
with finite quark chemical potential. 
Both the condensates are treated self-consistently 
by means of the gap equations. 
In section 3, the solutions of the gap equations are numerically given and 
the behaviors of the tensor condensates and the dynamical quark masses 
related to the light quarks ($u$ and $d$ quarks) and the strange quark 
are investigated. 
The last section is devoted to a summary and concluding remarks.

\section{Mean field approximation for the Nambu-Jona-Lasinio model with tensor-type four-point interaction between quarks}

Let us start from the three-flavor Nambu-Jona-Lasinio model with tensor-type \cite{IJMP,oursPTP} 
four-point interactions between quarks. 
The Lagrangian density can be expressed as 
\beq\label{2-1}
& &{\cal L}=\mathcal{L}_0+\mathcal{L}_m+\mathcal{L}_S+\mathcal{L}_{T}+\mathcal{L}_{D}\ , 
\nonumber\\
& &{\cal L}_0={\bar \psi}i\gamma^\mu \partial_{\mu}\psi , \nonumber \\
& &{\cal L}_m=-{\bar \psi} \vec m_0\psi , \nonumber \\
& &{\cal L}_S={G_s}\sum^8_{a=0}[({\bar \psi}\lambda_a\psi)^2+({\bar \psi}i\lambda_a\gamma_5\psi)^2],\nonumber \\
& &{\cal L}_T=-\frac{G_T}{4}\sum^8_{a=0}\left[({\bar \psi}\gamma^\mu\gamma^\nu\lambda_a\psi)
({\bar \psi}\gamma_\mu\gamma_\nu\lambda_a\psi)
+({\bar \psi}i\gamma_5\gamma^\mu\gamma^\nu\lambda_a\psi)({\bar \psi}i\gamma_5\gamma_\mu\gamma_\nu\lambda_a\psi)
\right] , \nonumber \\
& &{\cal L}_D=-G_D\left[\text{det}\bar\psi(1-\gamma_5)\psi + \text{det}\bar\psi(1+\gamma_5)\psi\right],
\eeq
where $\vec m_0$ represents a current quark mass matrix in flavor space 
as follows : 
\begin{align}\label{2-2}
	\vec m_0&= \text{diag}\left( m_u,m_d,m_s \right) \ .
\end{align}
Here, ${\cal L}_T$ represents a four-point tensor interaction between quarks in the three-flavor case 
which preserves chiral symmetry. 
Also, ${\cal L}_D$ represents so-called the Kobayashi-Maskawa-'t Hooft or the determinant 
interaction term
which leads to the six-point interaction between quarks in the three-flavor case. 
In this paper, we introduce minus sign in ${\cal L}_D$ in which we take $G_D>0$ from the beginning.

Hereafter, we treat the above model within the mean field approximation and 
ignore non-diagonal components of the condensates in a flavor space. 
Therefore, terms in the summation over $a$ 
are restricted to the diagonal entries 
with $a=0, 3$ and 8 
in ${\cal L}_S$ : 
\begin{align}\label{2-3}
	\sum^{8}_{a=0}[({\bar \psi}\lambda_a\Gamma\psi)^2] &
\longrightarrow
 \sum_{a=0,3,8}[({\bar \psi}\lambda_a\Gamma\psi)^2]\nonumber \\
	&\quad
=\frac{2}{3}\left[\left(\bar u \Gamma u +\bar d \Gamma d +\bar s\Gamma s\right)\right]^2 
	+\left[ \left(\bar u\Gamma u - \bar d\Gamma d\right)\right]^2 \nonumber\\
	& \qquad +\frac{1}{3}\left[\left(\bar u\Gamma u +\bar d\Gamma d -2\bar s\Gamma s\right)\right]^2 \nonumber \\
	&\quad
= 2(\bar u\Gamma u)^2 +2(\bar d\Gamma d)^2+2(\bar s\Gamma s)^2 \ .
\end{align}
Here, $\Gamma$ means products of any gamma matrices or unit matrix.
Also, in the determinant interaction term, ${\cal L}_D$, 
the same approximation is adopted, namely, the off-diagonal matrix elements in 
the flavor space are omitted: 
\begin{align}\label{2-4}
	&\text{det}\bar\psi\left(1-\gamma_5\right)\psi + \text{det}\bar\psi\left(1-\gamma_5\right)\psi \nonumber \\ 
	&\longrightarrow
 \text{det}
	\begin{pmatrix}
		\bar u(1-\gamma_5)u & 0 & 0 \\
		 0 & \bar d(1-\gamma_5)d &0 \\
		0 & 0 &\bar s(1-\gamma_5)s 
	\end{pmatrix}
\nonumber\\
&\qquad + \text{det}
	\begin{pmatrix}
		\bar u(1+\gamma_5)u & 0 & 0 \\
		0 & \bar d(1+\gamma_5)d &0 \\
		0 & 0 &\bar s(1+\gamma_5)s 
	\end{pmatrix} \nonumber\\
	&= 2(\bar uu) (\bar dd) (\bar ss) 
\nonumber\\
&\quad
+ 2(\bar uu) (\bar d\gamma_5d) (\bar s\gamma_5s)+2(\bar u\gamma_5u) (\bar dd) (\bar s\gamma_5s)+2(\bar u\gamma_5u) (\bar d\gamma_5d) (\bar ss) \ .
\end{align}
Secondly, in order to consider the spin polarization under the mean field approximation, 
the tensor condensate $\braket{{\bar q}\gamma^1\gamma^2 q}$ and 
$\braket{{\bar q}\gamma^2\gamma^1 q}$ are considered 
in ${\cal L}_T$ 
because 
$\gamma^1\gamma^2=i\Sigma_3$. 
Here,
\beq\label{2-5}
\Sigma_3=-i\gamma^1\gamma^2=
\begin{pmatrix}
		\sigma_3 & 0  \\
		0 & \sigma_3 
	\end{pmatrix} \ , 
\eeq
where $\sigma_3$ represents the third component of the Pauli matrix. 
Thus, we consider two tensor condensates under the mean field approximation as 
\beq\label{2-6}
& &F_3= -G_T\braket{{\bar \psi}\Sigma_3\lambda_3\psi}\ , \nonumber\\
& &F_8= -G_T\braket{{\bar \psi}\Sigma_3\lambda_8\psi}\ . 
\eeq
For each quark flavor, the tensor condensates are reexpressed as 
\beq\label{2-7}
& &F_u=F_3+\frac{1}{\sqrt{3}}F_8\ , \nonumber\\
& &F_d=-F_3+\frac{1}{\sqrt{3}}F_8\ , \nonumber\\
& &F_s=-\frac{2}{\sqrt{3}}F_8\ . 
\eeq
Of course, the chiral condensates $\langle {\bar q}q\rangle$ should be taken into account. 
We introduce the dynamical quark masses ${\cal M}_f$ 
without the determinant interaction term 
by using the chiral condensates as 
\beq\label{2-8}
& &{\cal M}_u=-4G_s\langle {\bar u}u\rangle\ , \nonumber\\
& &{\cal M}_d=-4G_s\langle {\bar d}d\rangle\ , \nonumber\\
& &{\cal M}_s=-4G_s\langle {\bar s}s\rangle\ . 
\eeq
These expressions are only valid if the mixing term is not considered.

Thus, under the mean field approximation, the Lagrangian density (\ref{2-1}) reduces to 
\begin{align}
\label{2-9}
\mathcal L_{MF}=&{\bar \psi}(i\gamma^\mu \partial_\mu -\vec M_q -\vec F \Sigma_3)\psi \nonumber \\
&-\sum_f\frac{{\cal M}_f^{2}}{8G_s}+\frac{F_3^2+F_8^2}{2G_T} -\frac{G_D}{16G_s^3}{\cal M}_u{\cal M}_d{\cal M}_s\ , 
\end{align}
where $f=u,\ d$ or $s$ and 
\begin{align}\label{2-10}
&\vec M_q =\text{diag.}\left(m_u + {\cal M}_u+\frac{G_D}{8G_s^2}{\cal M}_d{\cal M}_s\ ,
\right.
\nonumber\\
&\qquad\qquad\qquad \left. \ 
 m_d + {\cal M}_d 
+\frac{G_D}{8G_s^2}{\cal M}_s{\cal M}_u\ , 
\right.
\nonumber\\
&\qquad\qquad\qquad \left. \ m_s + {\cal M}_s +\frac{G_D}{8G_s^2}{\cal M}_u{\cal M}_d \right) \nonumber\\
&\qquad =\text{diag.}(M_u,\ M_d,\ M_s) \ , 
\end{align}
where ${\vec M}_q$ represents the constituent quark mass matrix with the flavor mixing due to the determinant interaction term.

Introducing the quark chemical potential $\mu$ in order to consider a quark matter 
at finite density, 
the Hamiltonian density can be obtained from the mean field Lagrangian density as
\beq\label{2-11}
{\cal H}_{MF}-\mu{\cal N}
&=&{\bar \psi}\left(-i{\mib \gamma}\cdot {\mib \nabla}+\vec M_q-\mu\gamma^0+\vec F\Sigma_3
\right)\psi
\nonumber\\
& &
+\sum_f\frac{{\cal M}_f^{2}}{8G_s}+\frac{F_3^2+F_8^2}{2G_T} +\frac{G_D}{16G_s^3}{\cal M}_u{\cal M}_d{\cal M}_s\ , 
\eeq
where ${\cal N}$ represents the quark number density, $\psi^{\dagger}\psi$.

Let us derive the effective potential or the thermodynamic potential at zero temperature. 
The Hamiltonian density (\ref{2-11}) 
can be rewritten as 
\beq
{\cal H}_{MF}-\mu{\cal N}
&=&\psi^{\dagger}(h_T-\mu)\psi 
+\sum_f\frac{{\cal M}_f^{2}}{8G_s}+\frac{F_3^2+F_8^2}{2G_T} +\frac{G_D}{16G_s^3}{\cal M}_u{\cal M}_d{\cal M}_s\ , 
\ \  \label{2-12}\\
h_T&=&-i \gamma^0{\mib\gamma}\cdot {\mib \nabla}+ \gamma^0\vec M_q+ \vec F\Sigma_3 .
\label{2-13} 
\eeq
In order to obtain the eigenvalues of the single-particle Hamiltonian $h_T$, namely the energy eigenvalues of single quark, it is necessary to diagonalize $h_T$, 
the eigenvalues of which can be obtained easily as 
\begin{align}
\label{2-14}
	E^f_{p_1,p_2,p_3,\eta} &=
\sqrt{p_3^2+\left(\sqrt{p_1^2+p_2^2+M_{f}^2}+\eta F_f\right)^2}\ ,
\end{align} 
where $\eta=\pm 1$.

Thus, we can easily evaluate the thermodynamic potential with the above single-particle energy eigenvalues. 
Then, the thermodynamic potential $\Phi$ can be expressed as
\begin{align}
\label{2-15}
	\Phi =
	&\sum_{f,\alpha,\eta} \int \frac{dp_3}{2\pi}\int \frac{dp_1}{2\pi}\int \frac{dp_2}{2\pi}  \left(E^f_{p_1,p_2,p_3,\eta}-\mu\right) \theta\left(\mu-E^f_{p_1,p_2,p_3,\eta}\right) \nonumber \\
		&-\sum_{f,\alpha,\eta} \int \frac{dp_z}{2\pi}\int \frac{dp_x}{2\pi}\int \frac{dp_y}{2\pi} E^f_{p_x,p_y,p_z,\eta} \nonumber \\
		&+ \sum_f\frac{{\cal M}_f^2}{8G_s} + \frac{F_3^2+F_8^2}{2G_T} + \frac{G_D}{16G_s^3}{\cal M}_u{\cal M}_d{\cal M}_s\ , 
\end{align}
where $\alpha$ represents the color degree which leads to numerical factor $N_c\ (=3)$.
Here, $\theta(x)$ represents the Heaviside step function. 
The first and second lines in (\ref{2-15}) represent the positive-energy contribution of quarks and the vacuum contribution, respectively.

To determine the chiral condensates or the dynamical quark masses ${\cal M}_f$ and the tensor condensates 
$F_3$ and $F_8$, the gap equation is demanded as 
\beq\label{2-16}
\frac{\partial \Phi}{\partial {\cal M}_u} = 
\frac{\partial \Phi}{\partial {\cal M}_d} = 
\frac{\partial \Phi}{\partial {\cal M}_s} = 
\frac{\partial \Phi}{\partial {F_3}} = 
\frac{\partial \Phi}{\partial {F_8}} =0 \ .
\eeq

\section{Numerical results}

In this section, let us derive numerical results and give discussions about the effects of the flavor mixing .

%
\begin{table}[b]
\caption{Parameter sets of 3-flavor NJL model without/with the determinant interaction (Model O/Model I).}
\label{table:modelparameters}
\begin{center}
\begin{tabular}{c||c|c|c|c|c|c|c}
\hline
 & $m_u$& $m_d$& $m_s$ & $G_s$ & $G_T$ & $G_D$ & $\Lambda$\\
 & [/GeV]& [/GeV] & [/GeV] & [/GeV$^{-2}$] & [/GeV$^{-2}$] &  [/GeV$^{-5}$]
&[/GeV]
\\ \hline
Model O & 0.0055 & 0.0055 & 0.1375 & 5.5 & $2G_s$\ (=11.0) & 0 & 0.6314 \\
\hline
Model I & 0.0055 & 0.0055 & 0.1375 & 4.6 & $2G_s$\ (=9.2) &9.288/$\Lambda^5$ & 0.6314 \\
\hline
\end{tabular}
\end{center}
\end{table}

First, we summarize the parameter sets with/without the determinant interaction which we call 
Model I/Model O, respectively. 
When we calculate the thermodynamic potential in (\ref{2-15}), a
regularization scheme is necessary 
because   the vacuum contribution in the second line in (\ref{2-15}) gives the divergent contribution. 
Here, we adopt the three-momentum cutoff scheme and introduce the three-momentum cutoff $\Lambda$. 
In Model O, which does not include the determinant interaction, the parameters are given so as to reproduce the 
pion decay constant and dynamical quark masses or pion mass and kaon mass. 
As for the tensor-type interaction, this interaction should be derived 
from a two-gluon exchange interaction in QCD \cite{oursPTEP4}. 
However, since the NJL model cannot be derived from the QCD Lagrangian directly, 
we  adopt $G_T$ as a free parameter in this model. 
If we assume that the tensor-type interaction term is derived by the Fierz transformation 
of the scalar-type four-point interaction term, 
$G({\bar \psi}\psi)^2$, in the NJL model, we can obtain the relationship $G_T=2G_S$. 
On the other hand, the value of $G_T$ can be determined by the vacuum properties of pion and $\rho$ meson as 
in Ref.\citen{Jaminon:2002}.  
Since, in the following, we discuss the system at finite density, this treatment may not be adequate. 
Thus, we treat $G_T$ as a free parameter. 
Similarly, in Model I in which the determinant interaction $G_D$ is introduced, 
the model parameters are determined so as to reproduce the masses of eta and eta prime mesons 
adding to pion and kaon masses and the pion decay constant \cite{HK}.

\subsection{Case of no flavor mixing (Model O)}

\begin{figure}[t]
      \begin{tabular}{cc}
	\begin{minipage}[t]{0.45\hsize}
	\begin{center}
		\includegraphics[height=4cm]{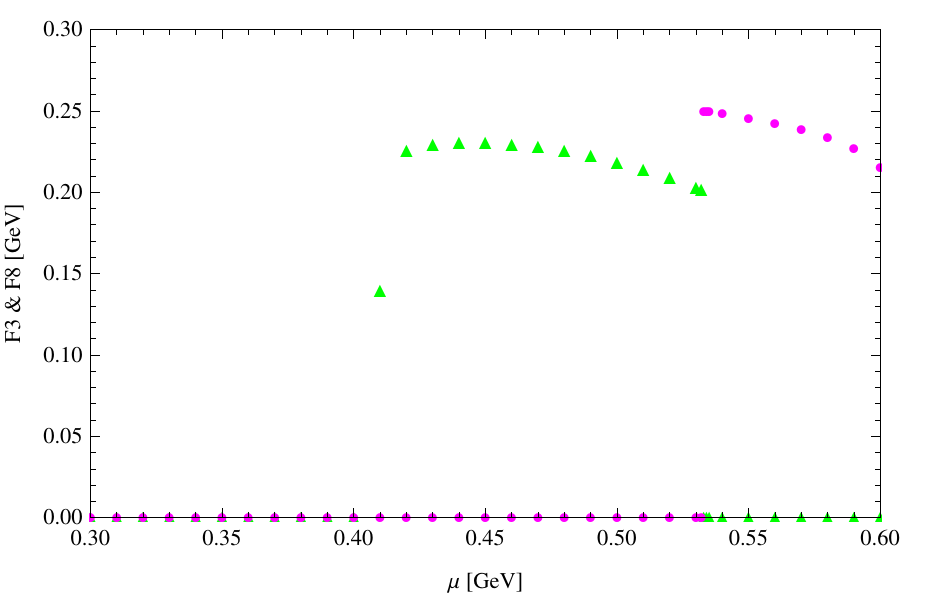}
	\caption{Tensor condensates $F_3$ and 
$|F_8|$ 
are depicted as a function of the quark chemical potential $\mu$. 
The triangle and circle represent $F_3$ and $|F_8|$, respectively. 
}
\label{fig:O38tensorcond}
	\end{center}
	\end{minipage}
\qquad
	\begin{minipage}[t]{0.45\hsize}
	\begin{center}
		\includegraphics[height=4cm]{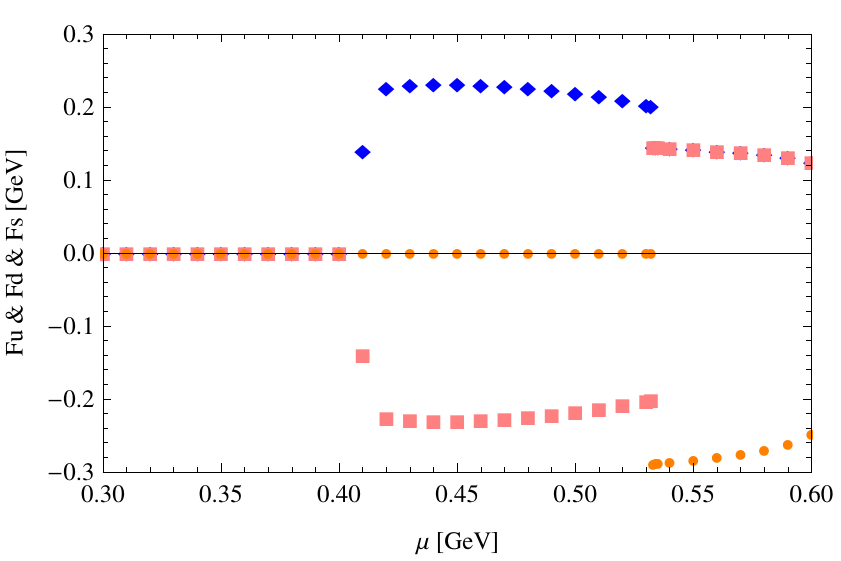}
	\caption{Tensor condensates for each flavor are depicted as a function of the quark chemical potential $\mu$. 
The diamond, squear and circle represent $F_u$, $F_d$ and $F_s$, respectively. 
}
\label{fig:OFtensorcond}
	\end{center}
	\end{minipage}
  \end{tabular}
\end{figure}

Figure \ref{fig:O38tensorcond} shows two possible tensor condensation $F_3$ and $F_8$ as a function of 
the quark chemical potential $\mu$ without the determinant interaction (Model O), namely 
the flavor mixing does not occur. 
In this figure, as for $F_8$, we plot $|F_8|$ because $F_8$ has a negative sign. 
In this case, first, at $\mu\approx 0.41$ GeV, $F_3$ appears, which is represented by green triangle in Fig.\ref{fig:O38tensorcond}. 
At $\mu\approx 0.532$ GeV, $F_3$ suddenly disappears. 
Instead of $F_3$, the condensate $F_8$ appears from $\mu\approx 0.532$ GeV and above it, which is 
represented by magenta circle in Fig.\ref{fig:O38tensorcond}. 
Thus, we conclude that two tensor condensates $F_3$ and $F_8$ does not 
coexist. 
Therefore, the relation $F_8\approx F_3/\sqrt{3}$ should not be demanded.
In Fig.\ref{fig:OFtensorcond}, the tensor condensates for each flavor, $F_u$ (blue diamond), $F_d$ (pink square) and 
$F_s$ (orange circle) are depicted. 
As is seen in Fig.\ref{fig:OFtensorcond}, $F_u=-F_d$ is satisfied from $\mu\approx 0.41$ to 0.532 GeV in which 
only $F_3$ appears. 
Above $\mu\approx 0.532$ GeV, $F_u=F_d$ and $F_s=2F_u$ are satisfied.

\begin{figure}[b]
	\begin{center}
		\includegraphics[height=4cm]{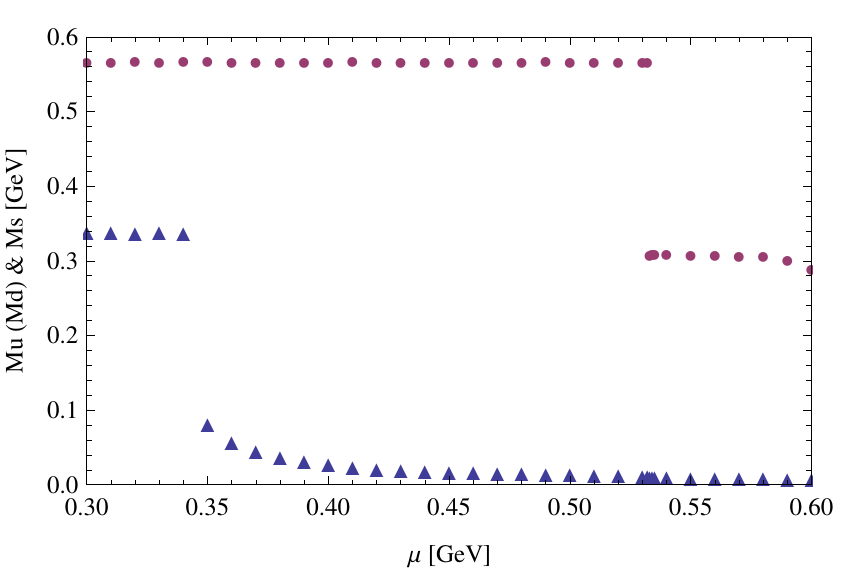}
	\caption{Constituent quark masses $M_u=M_d$ (blue triangle) and $M_s$ (violet circle) 
are depicted as a function of the quark chemical potential $\mu$. 
}
\label{fig:Omass}
	\end{center}
\end{figure}

\begin{figure}[t]
	\begin{center}
		\includegraphics[height=9cm]{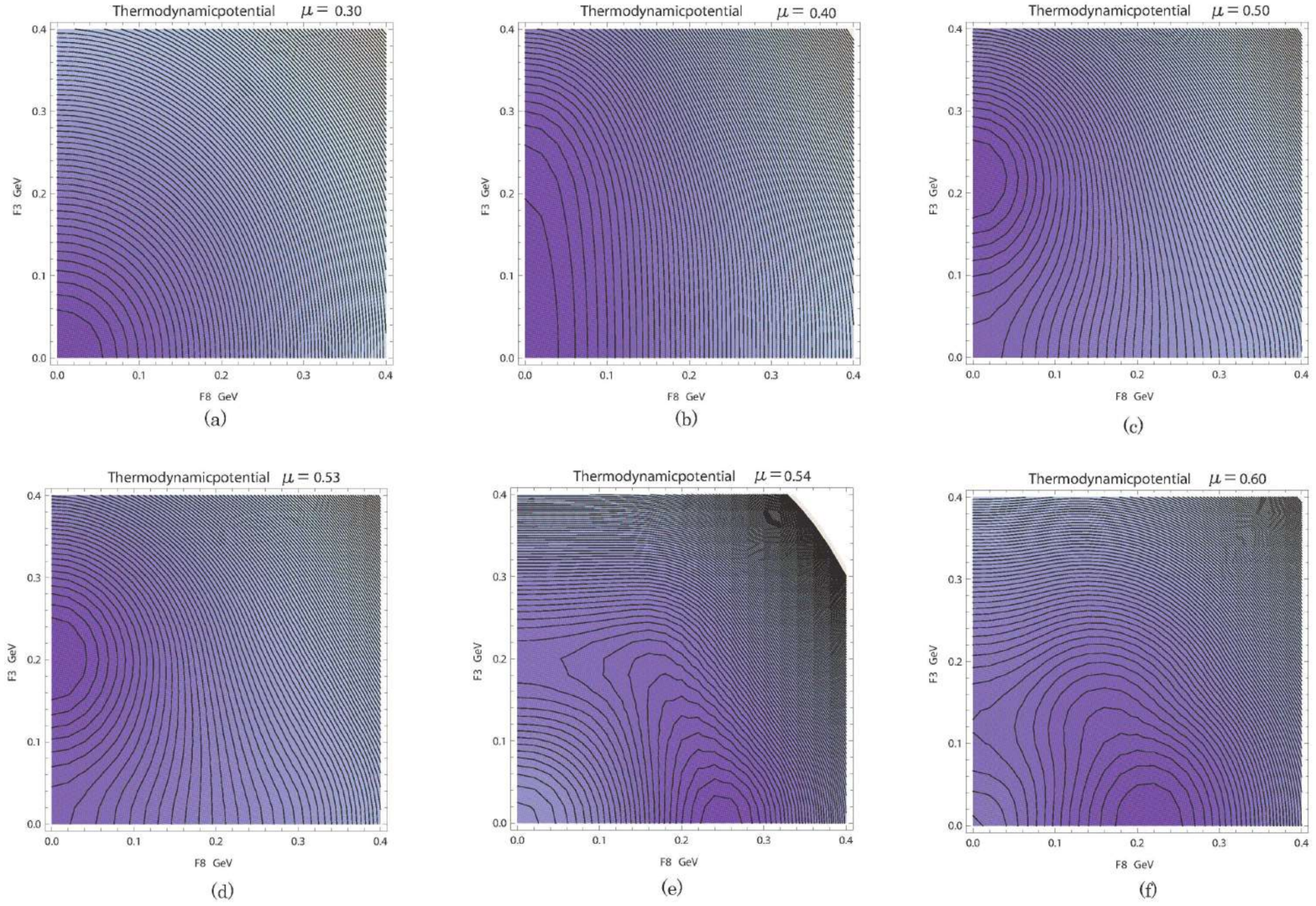}
	\caption{Contour plots of the thermodynamic potentials are depicted as (a) $\mu=0.30$ GeV, 
 (b) $\mu=0.40$ GeV,  (c) $\mu=0.50$ GeV,  (d) $\mu=0.53$ GeV,  (e) $\mu=0.54$ GeV and  (f) $\mu=0.60$ GeV. 
The horizontal and the vertical axies represent 
$F_8$ and $F_3$, respectively. 
}
\label{fig:Ocontour}
	\end{center}
\end{figure}

Figure \ref{fig:Omass} shows the constituent quark masses for $M_u\ (=M_d)$ (blue triangle) and $M_s$ (violet circle), 
respectively. 
At $\mu\approx 0.34$ GeV, the light quark masses decrease. 
At $\mu\approx 0.53$ GeV, the strange quark mass decreases, but above $\mu\approx 0.53$ GeV, 
the strange quark mass is not so changed. 
If $F_8$ is set to 0, the strange quark mass decreases above $\mu\approx 0.55$ GeV. 
Thus, it is shown that the tensor condensate with respect to the strangeness, $F_s$ leads to an almost constant 
strange quark mass and 
to 
the delay of the restoration of chiral symmetry for strangeness sector.

Figure \ref{fig:Ocontour} shows the contour plots of the thermodynamic potentials 
as   (a) $\mu=0.30$ GeV, 
 (b) $\mu=0.40$ GeV,  (c) $\mu=0.50$ GeV,  (d) $\mu=0.53$ GeV,  (e) $\mu=0.54$ GeV and  (f) $\mu=0.60$ GeV, 
respectively. 
The thermodynamic potential becomes lower as the color is darker in Fig.\ref{fig:Ocontour}. 
At $\mu=0.30$ GeV, the tensor condensates does not appear, $F_3=F_8=0$. 
Around $\mu=0.40$ GeV, $F_3$ begins to appear. 
From Fig.\ref{fig:Ocontour} (b), (c) and (d), the region from $\mu \approx 0.4$ to 0.53 GeV, 
the tensor condensate $F_3$ is non zero for the equilibrium configuration. 
However, at $\mu \approx 0.54$ GeV, $F_3$ disappears and $F_8$ appears.

\subsection{Case of flavor mixing (Model I)}

\begin{figure}[t]
      \begin{tabular}{cc}
	\begin{minipage}[t]{0.45\hsize}
	\begin{center}
		\includegraphics[height=4cm]{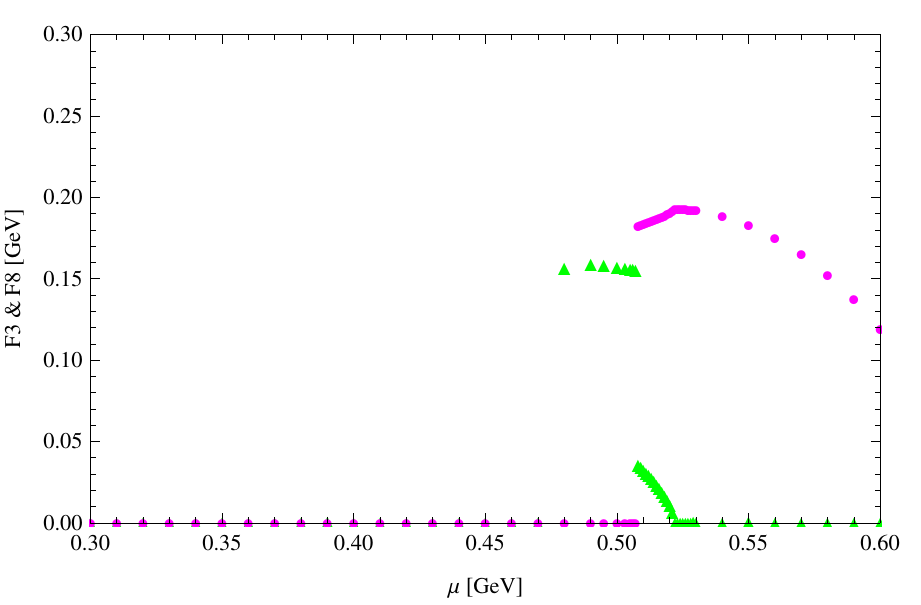}
	\caption{Tensor condensates $F_3$ and 
$F_8$ 
are depicted as a function of the quark chemical potential $\mu$. 
The triangle and circle represent $F_3$ and $F_8$, respectively.
}
\label{fig:I38tensorcond}
	\end{center}
	\end{minipage}
\qquad
	\begin{minipage}[t]{0.45\hsize}
	\begin{center}
		\includegraphics[height=4cm]{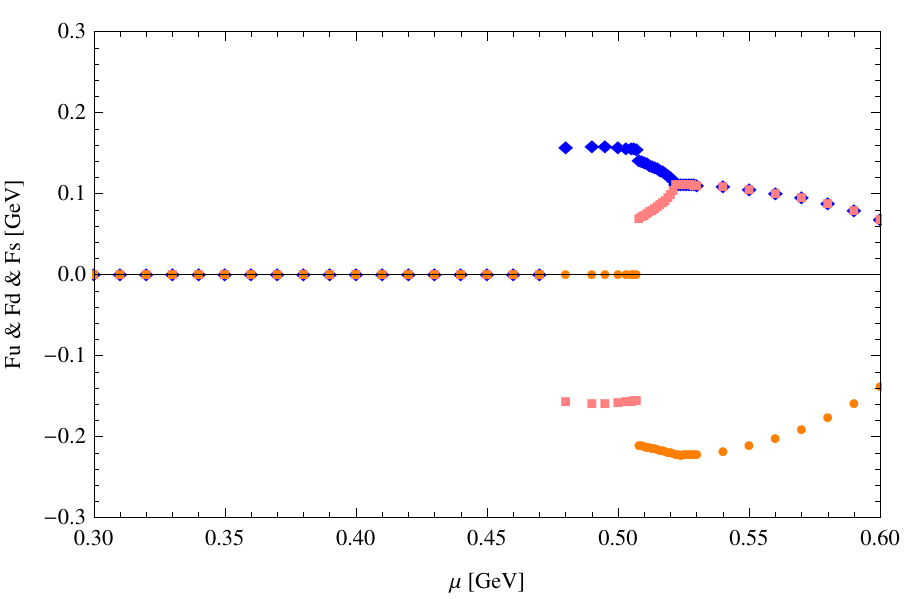}
	\caption{Tensor condensates for each flavor are depicted as a function of the quark chemical potential $\mu$. 
The diamond, square and circle represent $F_u$, $F_d$ and $F_s$, respectively. 
Because $F_u=F_d$ is satisfied above $\mu\approx 0.52$ GeV, the symbols of diamond and square in this figure 
overlap.
}
\label{fig:IFtensorcond}
	\end{center}
	\end{minipage}
  \end{tabular}
\end{figure}

Next, we discuss the numerical results with flavor mixing due to the determinant interaction. 
This interaction leads to the important $U_A(1)$-anomaly and the eta and eta prime mesons are well 
described \cite{KM,determinant}.  

Figure \ref{fig:I38tensorcond} shows two possible tensor condensation $F_3$ and $F_8$ as a function of 
the quark chemical potential $\mu$ 
with 
the determinant interaction (Model I) which leads to the flavor mixing. 
We also plot $|F_8|$ because $F_8$ has a negative sign. 
In this case, first, 
at $\mu\approx 0.48$ 
, $F_3$ (green triangle ) appears. 
At $\mu\approx 0.508$ GeV, $F_8$ suddenly appears and until $\mu\approx 0.52$ GeV, 
$F_3$ and $F_8$ coexist. 
Above $\mu\approx 0.52$ GeV, $F_3$ disappears and only $F_8$ remains. 
In Fig.\ref{fig:IFtensorcond}, the tensor condensates for each flavor, $F_u$ (blue diamond), $F_d$ (pink square) and 
$F_s$ (orange circle) are depicted. 
Below 
$\mu\approx 0.508$ GeV, $F_u=-F_d$ satisfies. 
However, in the region of $\mu\approx 0.508 \sim 0.52$ GeV, both the relations
$F_u=-F_d$ and $F_u=F_d$ are not valid. 
Above 
$\mu\approx 0.52$ GeV, $F_u=F_d$ and $F_s=2F_u$ are 
satisfied 
now. 
This behavior is originated from the flavor mixing, while this behavior does not seen in the no flavor mixing case.

\begin{figure}[b]
	\begin{center}
		\includegraphics[height=4cm]{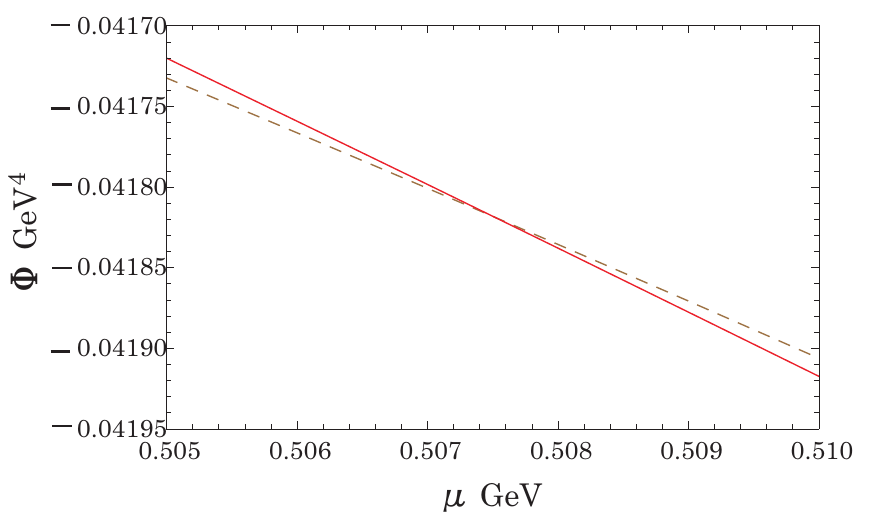}
	\caption{The thermodynamic potential 
$\Phi$ 
is depicted as a function of the quark chemical potential $\mu$. 
The dashed line represents the value of the thermodynamic potential with $F_3$ only.The solid line represents 
the case with $F_3\neq 0$ and $F_8\neq 0$, namely, the value of the thermodynamic potential in which 
two tensor condensates coexist.
}
\label{fig:Ipot}
	\end{center}
\end{figure}

The fact that two tensor condensates coexist is shown by the absolute value of the thermodynamic potential. 
Figure \ref{fig:Ipot} shows the value of the thermodynamic potential. 
The dashed line represents the value of the thermodynamic potential with $F_3$ only.
The solid line represents 
the case with $F_3\neq 0$ and $F_8\neq 0$, namely, the value of the thermodynamic potential in which 
two tensor condensates coexist.
Until $\mu\approx 0.507$ GeV, the thermodynamic potential with only $F_3\neq 0$ is lower than the 
coexistence case of $F_3\neq 0$ and $F_8\neq 0$. 
However, at 
$\mu\approx 0.5075$ 
GeV, the situation is reversed. 
Above $\mu\approx 0.508$ GeV, the coexistence of $F_3$ and $F_8$ is preferred. 

\begin{figure}[t]
	\begin{center}
		\includegraphics[height=4cm]{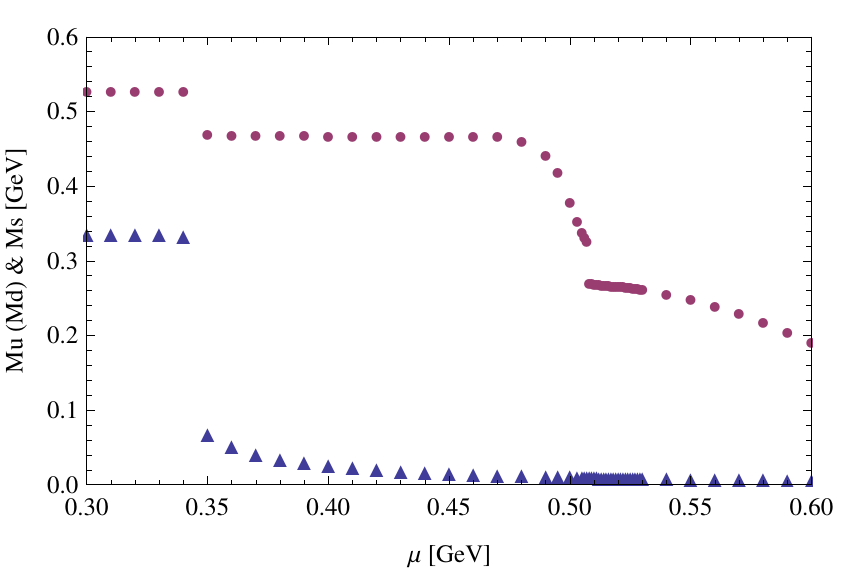}
	\caption{Constituent quark masses $M_u=M_d$ (blue triangle) and $M_s$ (violet circle) 
are depicted as a function of the quark chemical potential $\mu$. 
}
\label{fig:Imass}
	\end{center}
\end{figure}

\begin{figure}[b]
	\begin{center}
		\includegraphics[height=9cm]{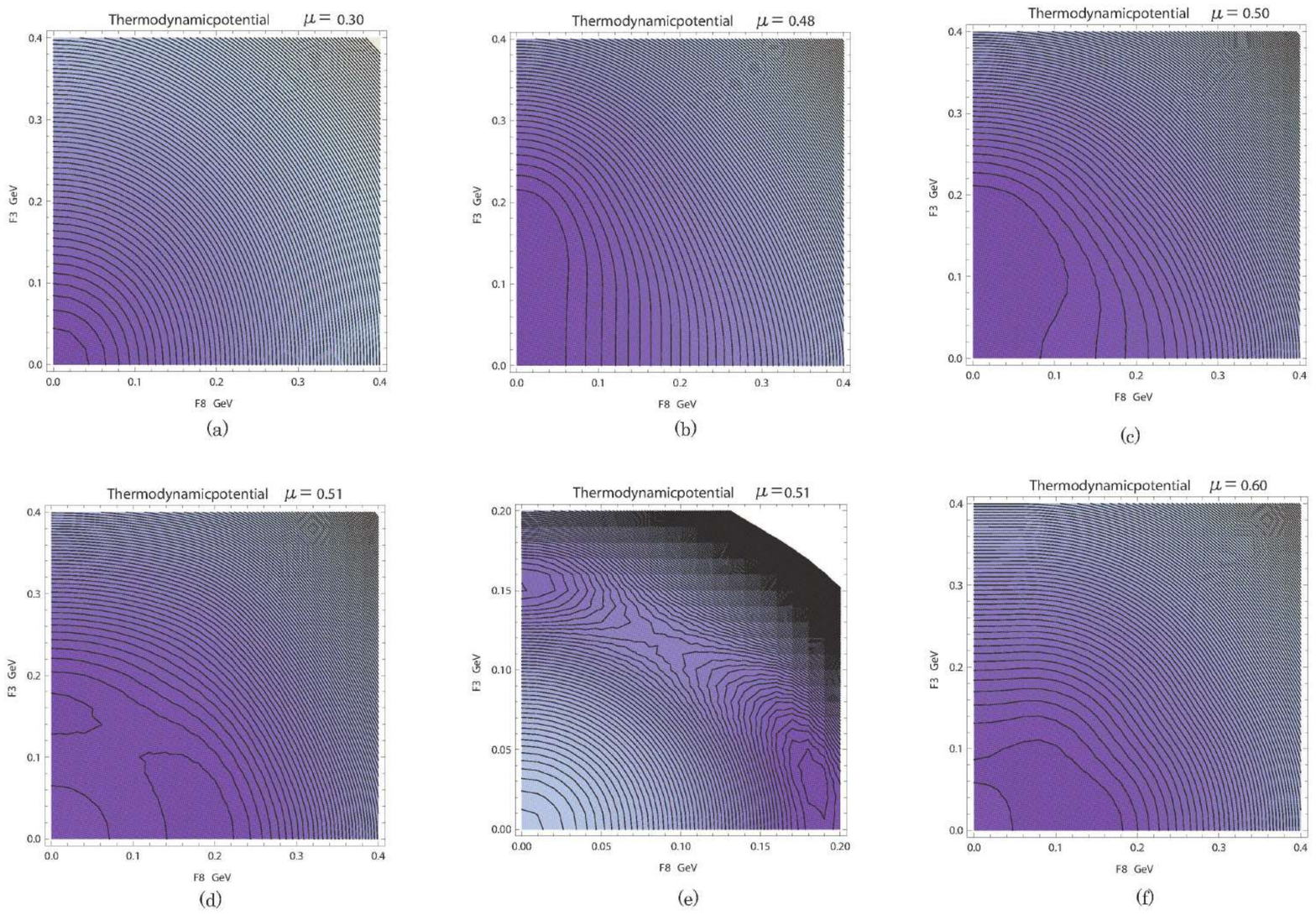}
	\caption{Contour plots of the thermodynamic potentials are depicted as (a) $\mu=0.30$ GeV, 
 (b) $\mu=0.40$ GeV,  (c) $\mu=0.50$ GeV,  (d) $\mu=0.53$ GeV,  (e) $\mu=0.54$ GeV and  (f) $\mu=0.60$ GeV. 
The horizontal and the vertical axies represent 
$F_8$ and $F_3$, respectively. 
}
\label{fig:Icontour}
	\end{center}
\end{figure}

Figure \ref{fig:Imass} shows the constituent quark masses for $M_u\ (=M_d)$ (blue triangle) and $M_s$ (violet circle), 
respectively. 
At $\mu\approx 0.34$ GeV, the light quark masses decrease. 
This behavior is the same as the result of no flavor mixing case.
At $\mu\approx 0.508$ GeV, the strange quark mass has a small gap and 
the light quark mass has a fine structure. 
This is due to the flavor mixing through the constituent quark masses $M_f$. 
Above $\mu\approx 0.508$ GeV, the strange quark mass gradually decrease. 
This is originated from the appearance of the tensor condensate $F_8$, namely $F_s$, 
as is seen in Fig.\ref{fig:Omass}. 
Thus, this behavior is not originated from the effect of the flavor mixing. .

Figure \ref{fig:Icontour} shows the contour plots of the thermodynamic potentials 
as (a) $\mu=0.30$ GeV, 
 (b) $\mu=0.48$ GeV,  (c) $\mu=0.50$ GeV,  (d) $\mu=0.51$ GeV,  (e) $\mu=0.51$ GeV same as (d) except for 
the scale and  (f) $\mu=0.60$ GeV, 
respectively. 
The thermodynamic potential decreases as the color becomes darker in Fig.\ref{fig:Icontour}. 
At $\mu=0.30$ GeV, the tensor condensates do not appear, $F_3=F_8=0$. 
Around $\mu=0.48$ GeV, $F_3$ begins to appear. 
At $\mu\approx 0.50$ GeV, $F_3$ has a finite value while $F_8=0$. 
However, at $\mu\approx 0.51$ GeV, $F_8$ begins to take a finite value adding to $F_3$. 
The detailed behavior at $\mu=0.51$ GeV is shown in Fig.\ref{fig:Icontour} (e). 
In (e), two local minima are seen. One is at the position with $F_3\approx 0.15$ GeV and $F_8=0$ and 
the other is at  $F_3\approx 0.02$ GeV and $F_8\approx 0.19$ GeV. 
As is indicated in Fig.\ref{fig:Ipot}, the right local minimum with $F_3\neq 0$ and $F_8\neq 0$ 
is a true minimum of the thermodynamic potential. 
Thus, there exists a region where the two tensor condensates coexist
due to the effect of the flavor mixing.

\section{Summary and concluding remarks}

The  possible formation of tensor condensates in three-flavor cold quark matter has 
been investigated taking  the Nambu-Jona-Lasinio model with the tensor-type four-point interaction 
as an effective model of QCD. 
In the three flavor case, it is necessary to consider the $U_A(1)$ anomaly which is incorporated 
in the Kobayashi-Maskawa-'t Hooft interaction or so-called determinant interaction with six-point interaction 
between quarks.
In our previous work \cite{oursPTEP2}, the tensor condensates  has been
investigated 
against 
the  color-flavor locked state which leads to a
color superconducting phase. 
In that paper, we showed that two tensor condensates may appear which
we denote by $F_3$ and $F_8$. 
Besides, we have treated only the three-flavor system in the chiral limit, namely all current quark masses being zero. 
As a result, the two condensates were related with each other, namely $F_8=F_3/\sqrt{3}$, when all the current quark masses are 
the same. 
In Ref.\citen{India}, the tensor condensates has also been considered
within the three-flavor NJL model with tensor interaction and 
the determinant interaction. 
However, almost all results have been obtained under the ansatz of  $F_8=F_3/\sqrt{3}$. 
In the present study, we did not consider this last approximation. 
Instead, 
we have estimated both tensor
condensates independently within the three-flavor NJL model 
with tensor interaction. 
In both the cases 
of the 
flavor mixing due to the determinant interaction and no flavor mixing, 
we have reinvestigated the cold quark matter with the finite quark chemical potential. 
We have  confirmed that the ansatz of  $F_8=F_3/\sqrt{3}$ is not valid except 
for the case of 
chiral limit.

Focusing on the determinant interaction in the three-flavor NJL model, which leads to the 
quark-flavor mixing, 
we have investigated the effect of flavor mixing  on the two types of the tensor condensates and 
dynamical quark masses.
As a result, the quantities related to the strange quark are 
affected by the determinant interaction, especially the behavior of the 
dynamical quark mass as a function of the quark chemical potential, 
while the quantities related to the light quarks are hardly affected. 
The tensor condensate related to the strange quark occurs at a rather small 
quark chemical potential compared with the case of no flavor mixing, namely, the case 
without the determinant interaction. 
The different behavior of the quark masses, which depend strongly on the presence of the 
determinant interaction, is the cause of this result. 
These behaviors are the same as the ones seen in the pseudovector condensates and the dynamical masses 
with pseudovector 
interaction 
and the flavor mixing \cite{Morimoto2}.
Under the model parameters used in this paper, 
the tensor condensate for light quarks 
and one for the strange quark coexist in a certain region of the quark chemical potential. 
On the other hand, if the flavor mixing is switched off, the coexistence region disappears. 
Thus, it is regarded as the effect of the flavor mixing.

The tensor condensates may lead to quark spin polarization. 
Therefore, to clarify the magnetic properties such as the spontaneous magnetization and 
magnetic susceptibility, it is interesting to investigate this problem
in future studies.
Further, the implication to the compact stars such as 
neutron stars and magnetars should be investigated by assuming the existence of the 
tensor condensates, related to the light quarks and the strange quark, while 
we have investigated the effect of the tensor condensate on the radius-mass relation of the hybrid quark star 
in the two-flavor case \cite{oursPR2}. 
This may be interesting future problem.

\section*{Acknowledgements}

Three of the authors (A.K, M.M and Y.T.) would like to express their sincere thanks to 
Professor K. Iida and Dr. E. Nakano for their helpful comments. 

\end{document}